\documentclass{article}
\usepackage{hiph-preprint}
\usepackage{graphics}
\usepackage{amssymb}
\usepackage{epsfig}

\volnumber{22} \issuenumber{1} \edyear{2005}                             %|
\frompage{000} \topage{000}                                              %|
\recrevdate{18 November, 2005}                                              %|
\def\rightmark{Testing the mechanism of QGP-induced energy loss}
\def\leftmark{I.~Vitev}
 
\title{Testing the mechanism of QGP-induced energy loss} 
\authors{ 
{Ivan Vitev
\index{Vitev, I.} % Abbreviated names of the author(s),
}\\[2.812mm]
{\normalsize
Los Alamos National Laboratory, 
Theoretical Division and Physics Division,\\ 
Los Alamos, NM 87545, USA\\[0.2ex] 
}}
 
\abstract{ We present an analytic model of jet quenching,
based on the (D)GLV energy loss formalism, to describe the
system size dependence of QGP-induced parton absorption in 
relativistic heavy ion collisions. Numerical simulations of 
the transverse momentum dependence of jet quenching are 
given for central Au+Au and Cu+Cu reactions. 
Low $p_T$ dijet correlations are shown to be sensitive 
to the reappearance of the lost energy as soft hadrons.
At high $p_T$ we find that the attenuation of dihadrons is 
similar to that of single inclusive particles.
Comparison to recent data from PHENIX and STAR
is given as a test of the jet quenching theory. }


 
\makeindex
\begin{document}
 
\maketitle

\section{Introduction}\label{intro}

In this contribution key results from the many-body QCD dynamics of
partons in dense nuclear matter are summarized. The emphasis is on
the systematic understanding of the signatures of final state radiative 
energy loss effects~\cite{eloss,Vitev:2002pf} in the quark-gluon 
plasma (QGP). Theoretical predictions are confronted by the  
recent RHIC data.

Tomographic determination of the properties of the medium
created in nucleus-nucleus collisions can  only be achieved 
through detailed numerical 
simulations~\cite{eloss,Vitev:2002pf,Wang:2005wv,Adil:2004cn,Vitev:2004gn}.
It is, however, useful to elucidate such studies with a 
simplified analytic model, which incorporates the essential 
features of jet quenching calculations~\cite{eloss}. 
We consider inclusive particle production, where the fragmentation
functions  $D_{h/_c}(z)$  are convoluted with the power law 
partonic cross  sections as follows
\begin{equation}
  \frac{d \sigma^{h}}{dyd^2p_T}  =  
\sum_c \int_{z_{\min} }^1 dz \; 
\frac{d \sigma^{c} (p_c = p_T/z) }{dyd^2p_{T_c}} \frac{1}{z^2} D_{h/_c}(z)
 \approx   \sum_c  \frac{A}{ p_{T_c}^n } \langle z \rangle^{(n-2)} 
D_{h/_c}(z)\;.
\label{hspectrum}
\end{equation}
In Eq.~(\ref{hspectrum})  $z_{\min} = p_T / p_{T_{c}\, \max }$ 
and $n$ is the partonic cross section power law index. 
The second input to the analytic model comes form
the Gyulassy-Levai-Vitev energy loss formalism~\cite{Gyulassy:1999zd},
extended by Djordjevic et al.  to  the  case  of  heavy 
quarks~\cite{Djordjevic:2003zk}. For the fractional
energy loss $\epsilon = { \Delta E }/{E}$  in (1+1)D Bjorken 
expansion  we find in the limit of large parton energy 
$2E/ \mu^2 L \gg 1$,
\begin{equation}
\frac{ \Delta E }{ E }
\approx \frac{ 9 C_R \pi \alpha_s^3  }{4}
 \frac{1}{A_\perp} \frac{dN^{g}}{dy} \, L 
 \; \frac{1}{E} \, \ln   \frac{2 E}{\mu^2 L}  +  \cdots  \; .
\label{analyt-de}
\end{equation}
The key to understanding the dependence of jet quenching on the 
nuclear species is the  $A$ or $N_{part}$ dependence of the 
characteristic  parameters in Eq.~(\ref{analyt-de}). We recall 
that ${dN^g}/{dy}  \propto  {dN^h}/{dy} \propto  A   \propto 
 N_{part}$,  $L \propto  A^{1/3}  \propto  N_{part}^{1/3} $, 
$ A_{\perp} \propto A^{2/3} \propto  N_{part}^{2/3}$.
Therefore, the  fractional energy loss, which is boost and gauge 
invariant when it enters physical observables, scales as   
$\epsilon = { \Delta E }/{E}  \propto  A^{2/3} 
\propto   N_{part}^{2/3}$.

\begin{figure}[t!]
\vspace*{0cm}
\includegraphics[width=2.2in,height=1.8in]{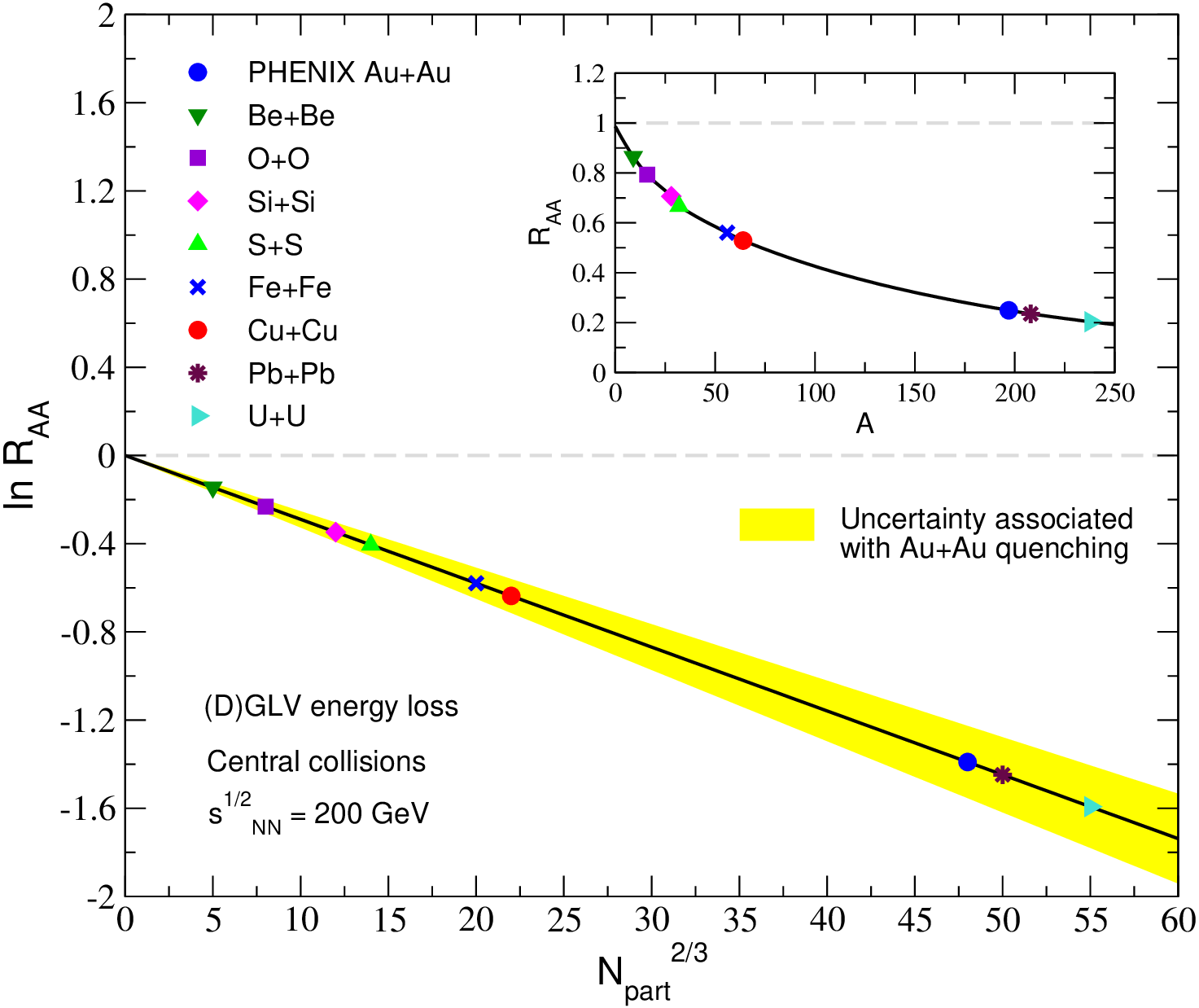}
\hspace*{0.2in}
\includegraphics[width=2.3in,height=1.8in]{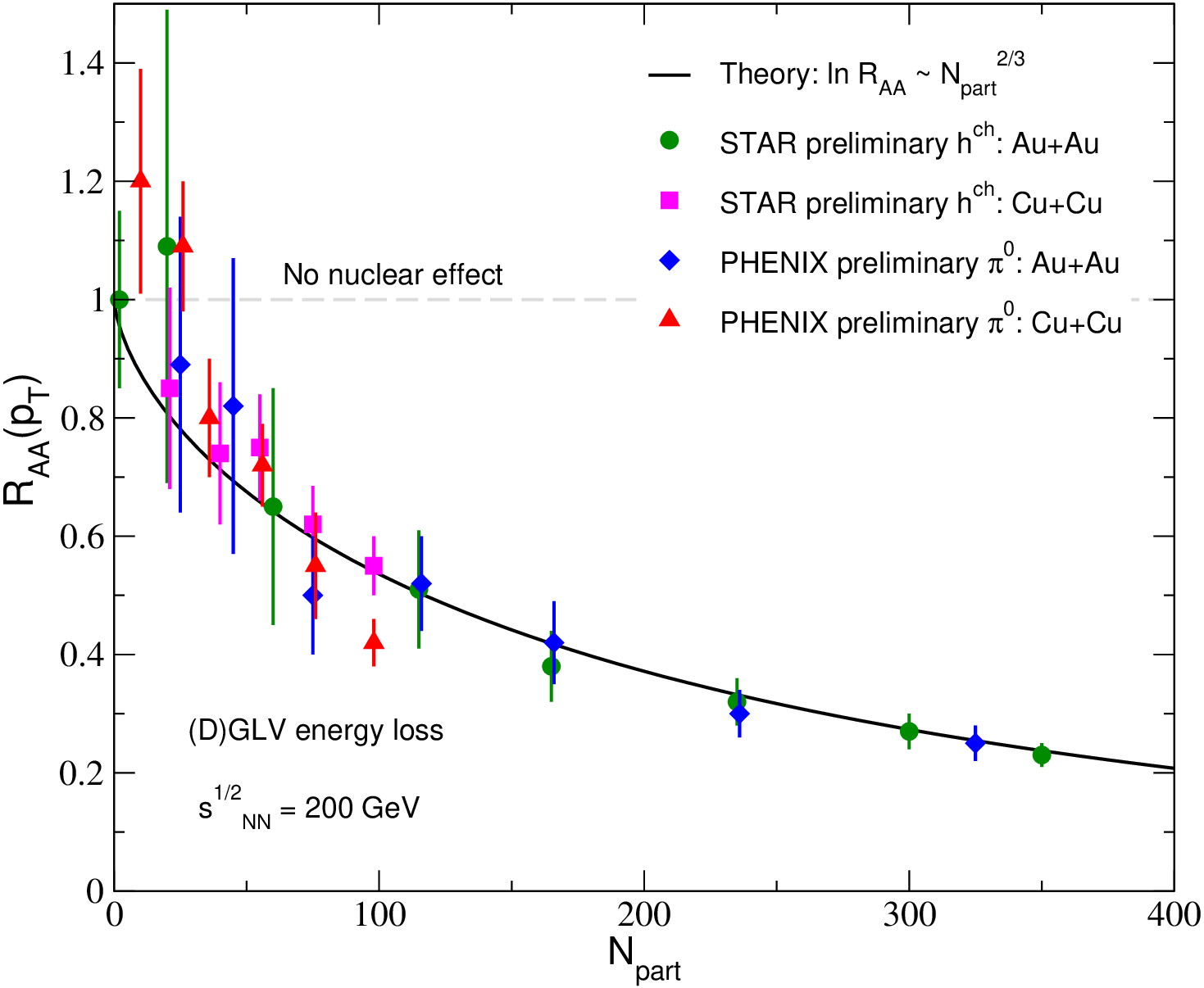}
\vspace*{-0.5cm}
\caption[]{ Left panel: the predicted linear relation between
$\ln R_{AA}$ and 
$N_{\rm part}^{2/3}$~\cite{eloss,Gyulassy:1999zd}.   
Quenching in head-on collisions is shown for
nuclear systems ranging form Be+Be to U+U. 
Right panel: comparison of the analytic model to the 
centrality dependent quenching in Cu+Cu and Au+Au 
reactions measured by STAR and 
PHENIX~\cite{Dunlop:2005xe,Shimomura:2005en}. }
\label{fig1}
\end{figure}

From Eqs.~(\ref{hspectrum}) and (\ref{analyt-de}) we 
can easily derive the system size dependence of the nuclear 
modification  factor
\begin{equation} 
R_{AB}  =  \frac{1}{N_{AB \;col}} \frac{ d\sigma^h_{AB}/dyd^2 p_T }
{ d \sigma^h  / dyd^2 p_T } 
 =  (1-\epsilon_{\rm eff})^{n-2}\;, 
\quad \ln R_{AA} = - \kappa N_{\rm part}^{2/3} \;, 
\label{raa-analyt} 
\end{equation} 
where $\kappa \approx (n-2)\epsilon_{\rm eff} / N_{\rm part}^{2/3}$.
The left panel of Fig.~1 shows the predicted $N_{\rm part}$
dependence of jet quenching at $\sqrt{s_{NN}}=200$~GeV. Central
collisions for systems ranging from  Be+Be to 
U+U are also shown.  The right panel of  Fig.~1 confronts
the theoretical model~\cite{eloss} with preliminary 
STAR~\cite{Dunlop:2005xe} and PHENIX~\cite{Shimomura:2005en} 
measurements in Au+Au and Cu+Cu collisions. Within the systematic
uncertainty, given by the difference in the measured $R_{AA}$ by the
two experiments, there is a good description of the centrality
dependence of jet quenching.

\section{Numerical results for the quenching of jets and 
dijets}\label{present}

Numerical evaluation of the nuclear modification factor $R_{AA}(p_T)$
in utrarelativistic heavy ion collisions is carried out as 
in~\cite{Vitev:2002pf,Vitev:2004gn}. The perturbative QCD hadron
production cross sections are modified by several, often 
competing, effects. Initial state multiple scattering 
leads to transverse momentum broadening 
of the incoming quarks and gluons and Cronin-like enhancement at 
intermediate transverse momenta. While nuclear shadowing is 
incorporated, its effects were found to be small, $|S(x,Q^2)-1| < 0.2$. 
In fact, dynamical nuclear enhanced power 
corrections  $\sim \xi^2A^{1/3}/Q^2$ may play
a dominant role in the modification of the DIS 
structure functions~\cite{Qiu:2003vd}. 
Their effect is negligible for $p_T > 5$~GeV hadron 
production~\cite{Qiu:2003vd}.

\begin{figure}[t!]
\vspace*{0cm}
\includegraphics[width=2.2in,height=1.8in]{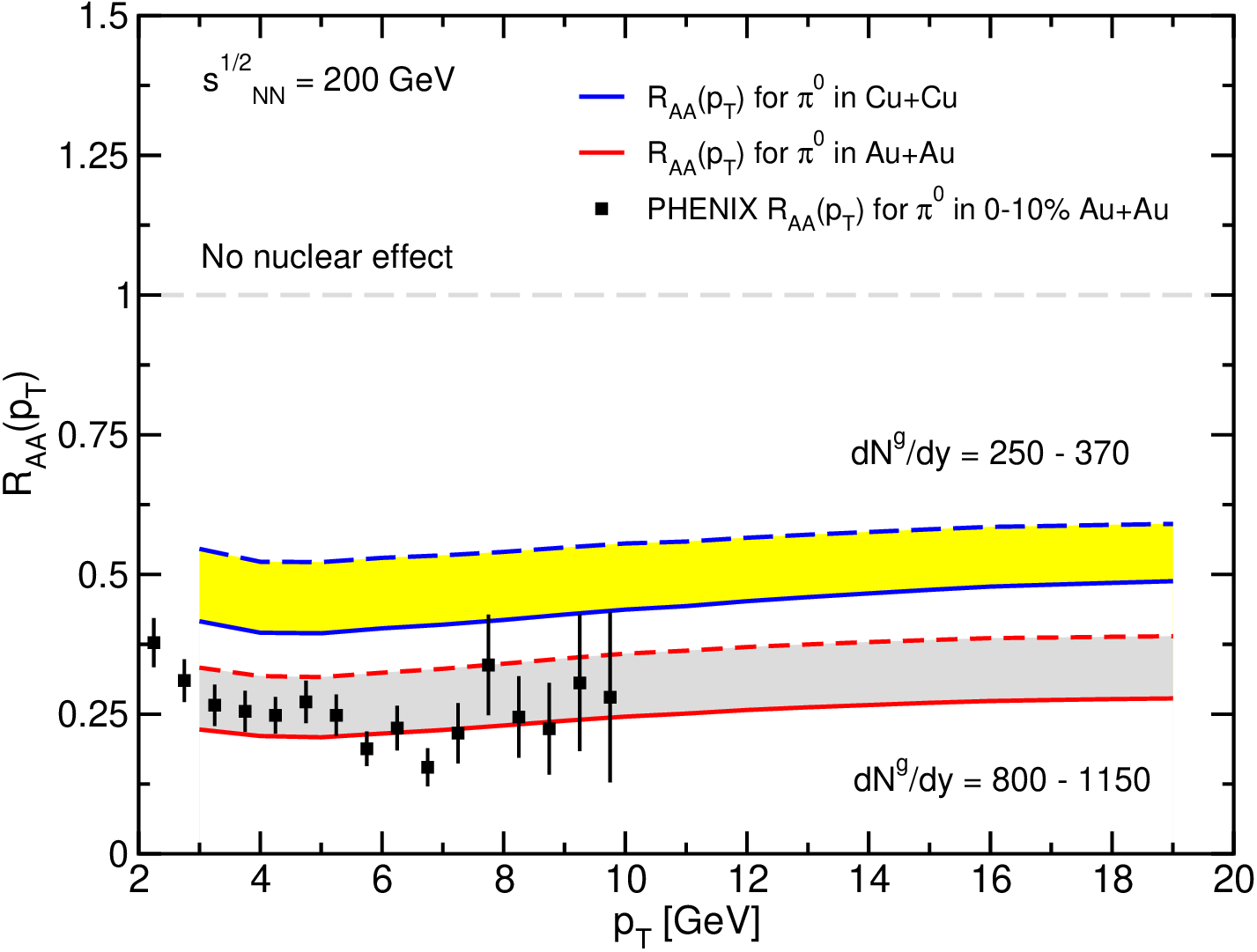}
\hspace*{0.2in}
\includegraphics[width=2.3in,height=1.8in]{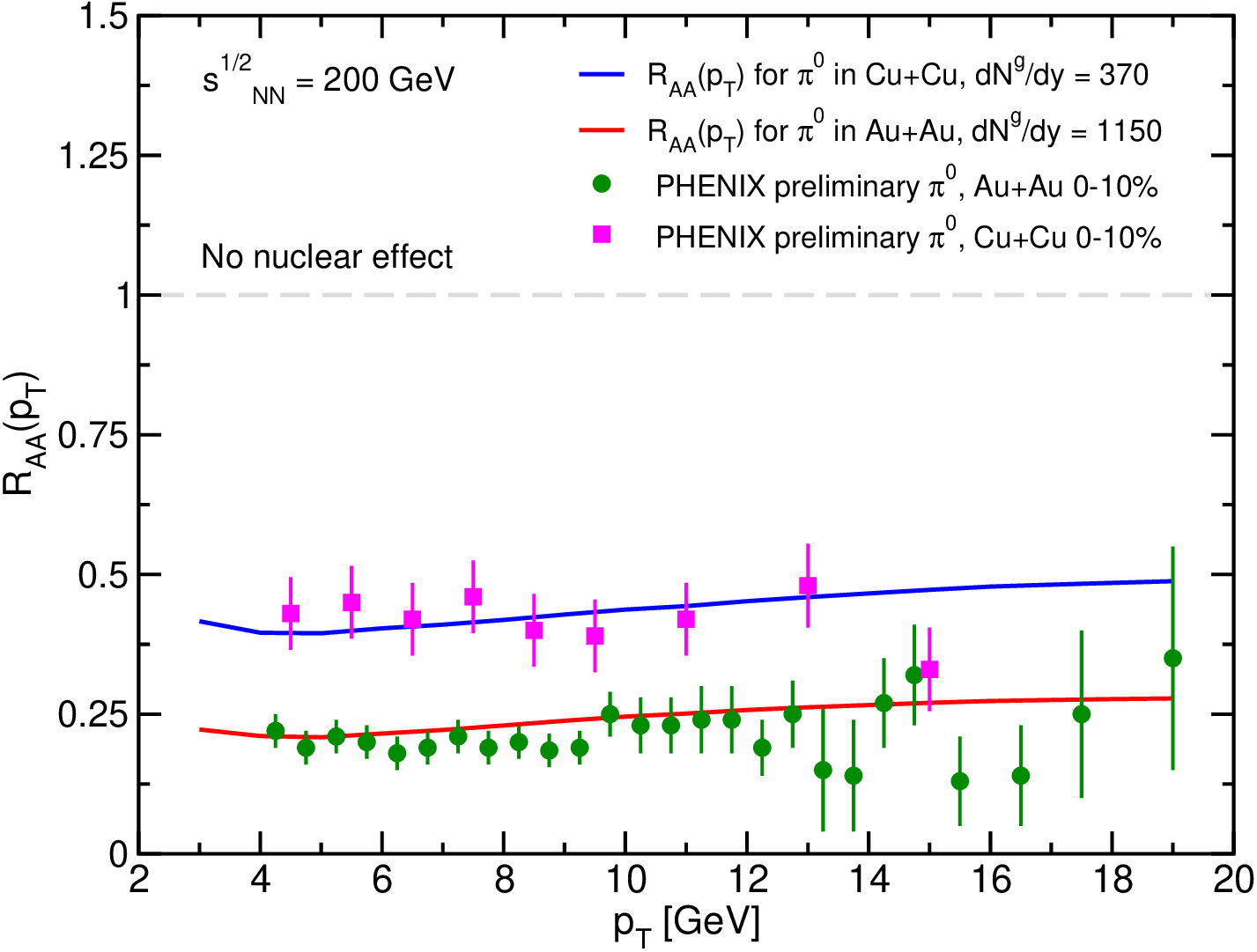}
\vspace*{-0.5cm}
\caption[]{Left panel: predicted quenching for single
inclusive $\pi^0$ ($\pi^+ + \pi^-$) production in central 
Au+Au and central Cu+Cu collisions to high 
$p_T = 20$~GeV~\cite{eloss,Vitev:2002pf}. Data is from 
PHENIX~\cite{Adler:2003qi}.
Right panel: comparison of the preliminary 0-10\% 
central Au+Au and Cu+Cu data from PHENIX~\cite{Shimomura:2005en}  
to the energy loss calculations with  
$dN^g/dy = 1150, 370$, respectively~\cite{eloss}. }
\label{fig2}
\end{figure}

When energy is lost in a deconfined medium of high 
color charge density $\rho \sim dN^g/dy /(\tau A_\perp)$ prior 
to fragmentation, two contributions arise from the 
medium-induced splitting of the hard parton.
With a suitable change of variables we find~\cite{Vitev:2005yg}
\begin{equation}
D_{h/c} (z)  \Rightarrow  \int_{0}^{1-z} d\epsilon \; P(\epsilon)  \; 
\frac{1}{1-\epsilon} D_{h/c} \left( \frac{z}{1-\epsilon} \right) 
+  \int_z^{1} d\epsilon \; \frac{dN^g}{d \epsilon}(\epsilon) \;  
 \frac{1}{ \epsilon }  D_{h/g} \left( \frac{z}{\epsilon} \right) \;\; .
\label{nucmod}
\end{equation}
Here, $z = p_{T}^h / p_{T_c} $ is the unmodified momentum 
fraction in the vacuum and $P(\epsilon)$ is the probability for
fractional energy loss $\epsilon$ due to multiple gluon 
emission~\cite{Vitev:2002pf,Adil:2004cn}. 
It is easy to verify the momentum sum rule 
for the hadronic fragments of the attenuated jet and the radiative gluons
\begin{equation}
  \sum_h \int_0^1 dz \; z  D_{h/c} (z)  =  1 - \langle \epsilon \rangle  
+  \langle \epsilon \rangle  = 1 \;.
\end{equation}

For the single inclusive particle  production at high $p_T$ 
the first term in Eq.~(\ref{nucmod}) dominates. The left 
panel of Fig.~2 shows the predicted transverse momentum 
dependence of $R_{AA}(p_T)$ for central Au+Au and Cu+Cu 
collisions at the top RHIC energy. The sensitivity of the 
calculation to the variation in the effective gluon rapidity 
density $dN^g/dy$ is illustrated. The quenching of jets is approximately 
$p_T$ independent in the range $ 5 \; {\rm GeV}   <  p_T < 20 \; {\rm GeV}$. 
Data in central Au+Au collisions is from PHENIX~\cite{Adler:2003qi}.
The left panel of Fig.~2 shows the comparison of the (D)GLV 
theory to the recent PHENIX data in both central Au+Au and 
central Cu+Cu collisions to much higher 
$p_T \sim 20$~GeV~\cite{Shimomura:2005en}.

\begin{figure}[t!]
\vspace*{0cm}
\includegraphics[width=2.6in,height=2.2in]{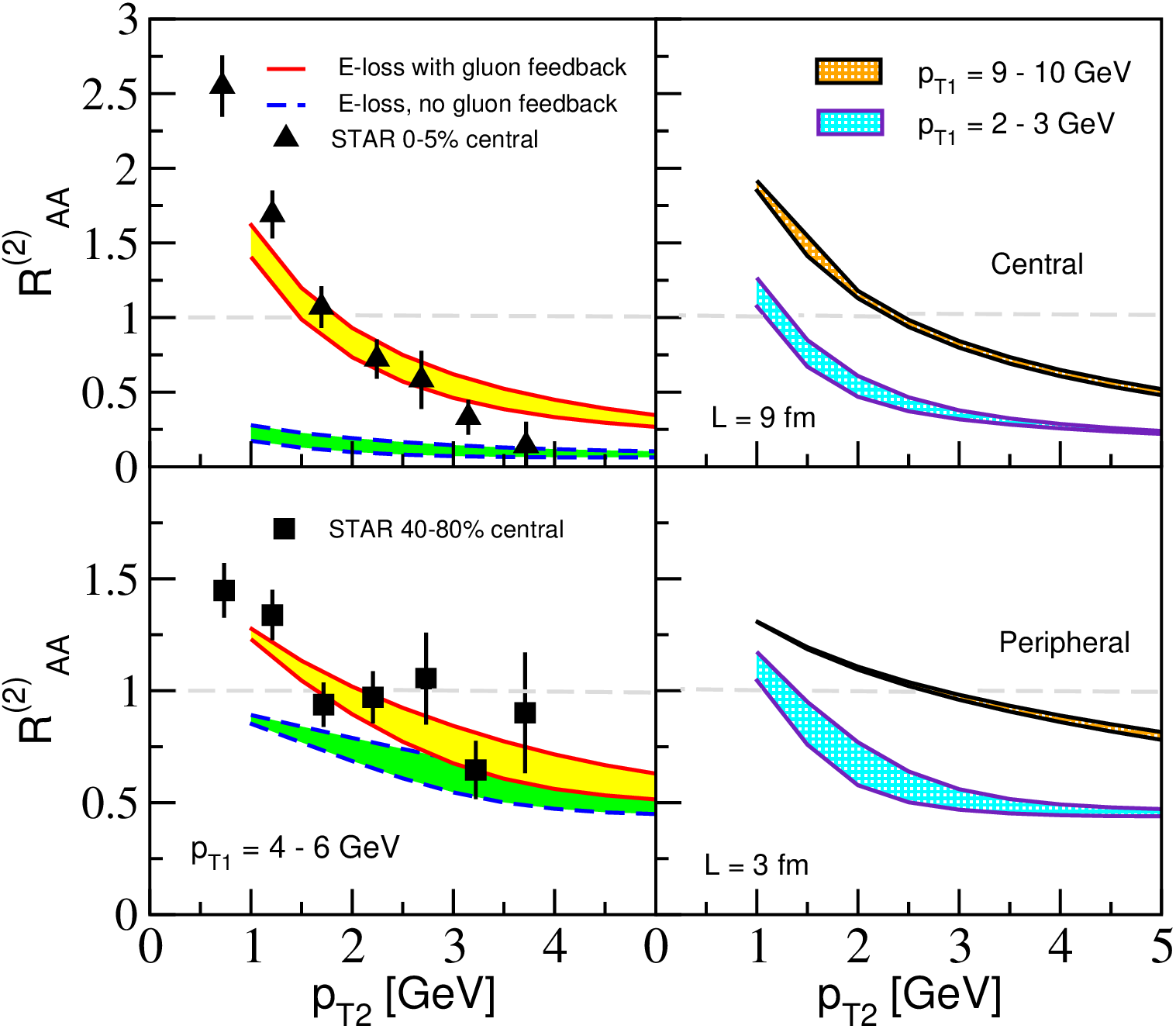}
\hspace*{0.2in}
\includegraphics[width=2.in,height=2.2in]{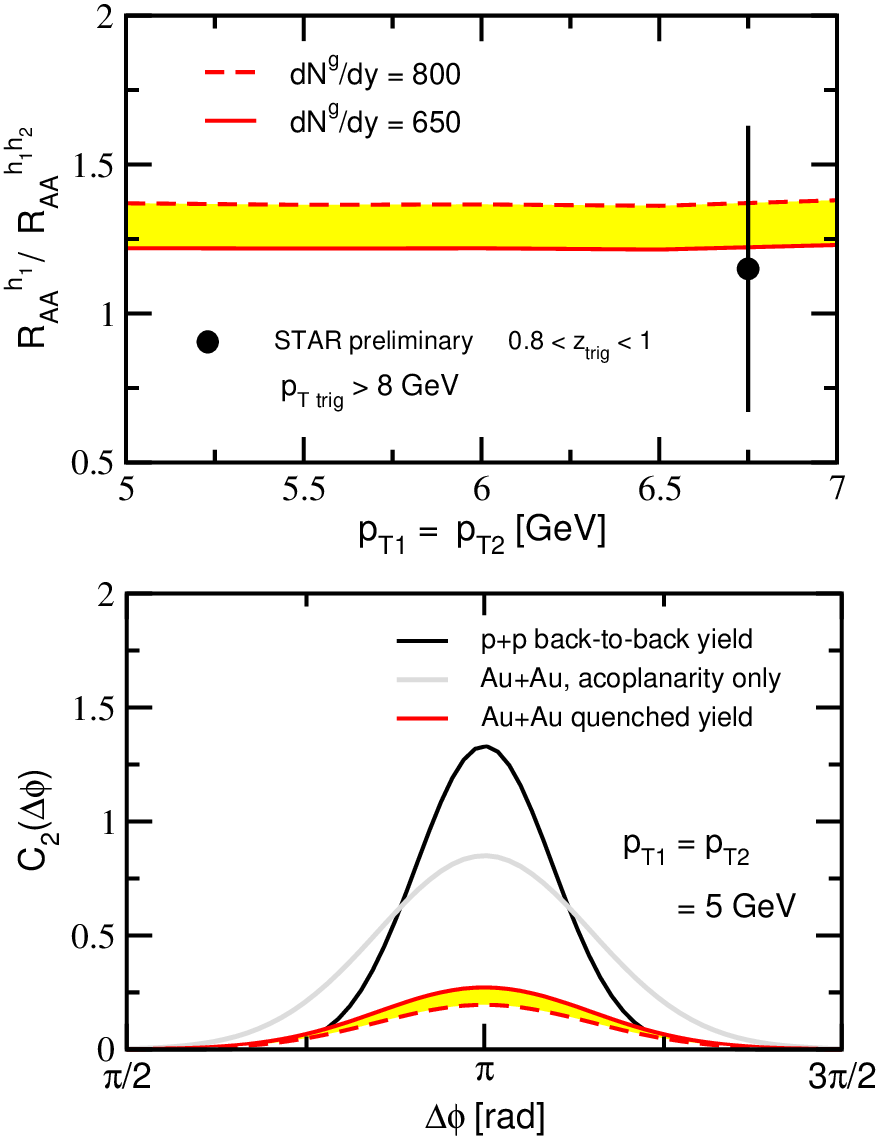}
\vspace*{-0.3cm}
\caption[]{Left panel: jet softening - redistribution of the 
energy from high $p_T$ to low $p_T$ particles~\cite{Vitev:2005yg}. 
Data is from STAR~\cite{Adler:2002tq}. 
Centrality and trigger momentum dependence of the 
nuclear dihadron modification are also shown.
Right panel: predicted ratio of the single inclusive
to double inclusive hadron suppression~\cite{Vitev:2004gn}. 
Preliminary data is from STAR~\cite{Magestro:2005vm}. }
\label{fig3}
\end{figure}

Qualitatively new information about the mechanisms of jet
modification in the QGP can be extracted from two-particle,
or dijet, correlations. In the medium-induced energy loss
scenario, the second term in  Eq.~(\ref{nucmod}) leads to
enhancement in the multiplicity of the away-side hadrons
associated with a high $p_T$ trigger~\cite{Vitev:2005yg}.
With $dN^g/d\epsilon$ calculated in the (D)GLV 
approach~\cite{Gyulassy:1999zd,Djordjevic:2003zk},
the redistribution of the energy is a parameter free
prediction~\cite{Vitev:2005yg}. The dependence  of the 
dihadron modification on the trigger and associated 
particle momenta, $p_{T_1}$ and $p_{T_2}$, and collision
centrality are shown in the left panels of Fig.~3.
Data is from STAR~\cite{Adler:2002tq}. Another aspect 
that high $p_T$ dijet correlations can  clarify 
is how opaque is the QGP. The right panel of Fig.~3  
compares the quenching of single inclusives to 
the quenching of dihadrons, 
$R_{AA}^{h_1} / R_{AA}^{h_1 h_2}$. Even without energy loss 
fluctuations and gluon feedback this ratio is $\sim 1.5$. 
For $\sqrt{s_{NN}}=62$~GeV $R_{AA}^{h_1} / R_{AA}^{h_1 h_2}
\sim 1.2 - 1.35$. 
Theoretical calculations are, thus, consistent with 
the preliminary STAR data~\cite{Magestro:2005vm} that 
improve upon earlier dijet measurements by being able to
recover high $p_T$ dijets. 
Note that for the data point in Fig.~3 $ z_{\rm trig} =  
p_{T \; {\rm assoc}} / p_{T \; {\rm trig}} $. 
Similar results were found by 
PHENIX~\cite{Magestro:2005vm}.

\section{Conclusions}\label{concl}

In this contribution we presented results from a 
theoretical study of the centrality and transverse
momentum dependence of jet quenching~\cite{eloss}.
At high $p_T$ and fixed $\sqrt{s_{NN}}$, in the 
absence of large additional partonic effects,
we find a universal dependence of $R_{AA}$ on 
$N_{\rm part}$~\cite{eloss}.  
The QGP-induced suppression at RHIC, independent of 
the system size, is approximately constant versus 
$p_T$~\cite{eloss,Vitev:2002pf}. Such transverse 
momentum behavior follows from the 
jet energy dependence of the (D)GLV energy 
loss~\cite{Gyulassy:1999zd,Djordjevic:2003zk} 
and its implementation in the factorized perturbative 
QCD  hadron production 
formalism~\cite{eloss,Vitev:2002pf,Vitev:2005yg}. 
Dihadron correlations are examined, within the same 
jet quenching approach, at  both low $p_T$ and high $p_T$. 
The redistribution of the lost energy in low $p_T$
particles is a verified, parameter free, prediction of the 
model~\cite{Vitev:2005yg}. At high transverse momenta, 
dijet suppression is shown to be only slightly larger
than the quenching of single inclusive particles~\cite{Vitev:2004gn}. 
The agreement with the recent STAR data~\cite{Dunlop:2005xe} 
indicates  that the QGP, created in Au+Au collisions at RHIC, 
is not fully opaque.

\vspace*{-.3cm}

\section*{Acknowledgments}
This research is supported in part by the US Department of Energy  
under Contract No. W-7405-ENG-3  and by 
the J. Robert Oppenheimer Fellowship of the Los Alamos National 
Laboratory.

\vfill\eject
\end{document}